\documentclass[pra,groupedaddress,superscriptaddress,reprint,twocolumn,showpacs,notitlepage]{revtex4-1}

\usepackage{graphicx,graphics,epstopdf,color,times,bm,bbm,dsfont,amssymb,amsmath,amsfonts,natbib}
\usepackage{amsmath}

\usepackage{times,txfonts}
\usepackage[hidelinks]{hyperref}
\hypersetup{
     colorlinks   = true,
     citecolor    = blue,
     linkcolor    = blue
}

\newcommand{\ignore}[1]{}

\let\oldsqrt\sqrt
\def\sqrt{\mathpalette\DHLhksqrt}
\def\DHLhksqrt#1#2{%
\setbox0=\hbox{$#1\oldsqrt{#2\,}$}\dimen0=\ht0
\advance\dimen0-0.2\ht0
\setbox2=\hbox{\vrule height\ht0 depth -\dimen0}%
{\box0\lower0.4pt\box2}}

\begin{document}
\title{Achieving sub-shot-noise sensing at finite temperatures}
\author{Mohammad Mehboudi}
\affiliation{Unitat de F\'isica Te\`orica: Informaci\'o i Fen\`omens Qu\`antics, Departament de F\'isica, Universitat Aut\`onoma de Barcelona, 08193 Bellaterra, Spain}
\author{Luis A. Correa}
\affiliation{Unitat de F\'isica Te\`orica: Informaci\'o i Fen\`omens Qu\`antics, Departament de F\'isica, Universitat Aut\`onoma de Barcelona, 08193 Bellaterra, Spain}
\affiliation{School of Mathematical Sciences, The University of Nottingham, University Park Campus, NG9 2RD Nottingham, United Kingdom}
\author{Anna Sanpera}
\affiliation{Unitat de F\'isica Te\`orica: Informaci\'o i Fen\`omens Qu\`antics, Departament de F\'isica, Universitat Aut\`onoma de Barcelona, 08193 Bellaterra, Spain}
\affiliation{Instituci\'o Catalana de Recerca i Estudis Avan\c{c}ats (ICREA), 08011 Barcelona, Spain} 
%
\begin{abstract}
We investigate sensing of magnetic fields using quantum spin chains at finite temperature and exploit quantum phase crossovers to improve metrological bounds on the estimation of the chain parameters. In particular, we analyze the $ XX $ spin chain and show that the magnetic sensitivity of this system is dictated by its adiabatic magnetic susceptibility, which scales extensively (linearly) in the number of spins $ N $. Next, we introduce an iterative feedforward protocol that actively exploits features of quantum phase crossovers to enable super-extensive scaling of the magnetic sensitivity. Moreover, we provide experimentally realistic observables to saturate the quantum metrological bounds. Finally, we also address magnetic sensing in the Heisenberg $ XY $ spin chain.  
\end{abstract}

\maketitle

%

\section{Introduction}

The field of \textit{metrology} is concerned with precise measurements of an unknown parameter and lies at the core of applications in cutting-edge time keeping, global positioning, or sensing of biological systems \cite{wynands2005atomic,lombardi2001time,kucsko2013nanometre}. In practice, technological advances are bringing the attainable measurement resolutions to a whole new level, as showcased, for instance, by the recent interferometric detection of gravitational waves \cite{abbott2016observation}. The active exploitation of quantum effects in high precision measurements, or \textit{quantum} metrology, holds promise for further improving the current metrological standards, which motivates an intense activity in this area of quantum technologies \cite{giovannetti2004quantum,giovannetti2011advances}.

The most generic metrological setting consists in coupling a probe to the parameter to be estimated through some suitable interaction. The outcomes of measurements performed on the probe are then used to build an estimate $ \lambda_\text{est} $ of the unknown parameter $ \lambda $. As a result of the central limit theorem, the corresponding statistical error in the estimation decreases as $\delta\lambda \sim 1/\sqrt{N} $ \cite{cramer1999mathematical}, where $ N $ is the number of repetitions or, equivalently, the number of \textit{uncorrelated} and independent probes used in the estimation. This type of scaling is often referred-to as shot noise or standard quantum limit \cite{giovannetti2004quantum}. On the contrary, if the probes are prepared in an \textit{entangled} state before locally interacting with the parameter, the statistical uncertainty can decrease, at most, as $ \delta\lambda\sim 1/N $, which is customarily termed \textit{Heisenberg scaling} \cite{PhysRevLett.96.010401}.   

In practice, it is very hard to beat the shot-noise scaling with independent probes \cite{PhysRevLett.79.3865}. This is primarily due to the fragility of entanglement to environmental noise \cite{demkowicz2012elusive}. In fact, few types of noise \cite{PhysRevLett.109.233601,PhysRevA.92.010102,smirne2015ultimate,PhysRevLett.111.120401,PhysRevX.5.031010,Yuan2016,PhysRevLett.115.170801,PhysRevA.84.012103} allow for better than shot-noise performance. Initializing of a large number of probes in pure entangled states on cue is also a demanding task, which poses an additional impediment to practical quantum metrology. It would be preferable to work instead with readily available preparations, such as (mixed) thermal states \cite{spedalieri2016thermal}. An additional problem arises from the fact that the optimal measurements to be performed on the $ N $ probes, i.e. those that minimize the uncertainty in the estimation, are often highly non-local collective measurements and thus, practically infeasible. 

One way around the inherent difficulties of sub-shot-noise parameter estimation is to consider \textit{interacting} probes rather than independent ones, so as to harness the metrological power of quantum many-body systems. With the advent of quantum simulators based on \textit{ultracold} atoms and ions, several paradigmatic Hamiltonians representing simple spin models are being implemented in a very controllable manner \cite{paredes2004tonks,lewenstein2007ultracold,simon2011quantum,Toskovic2016}, which paves the way towards practical quantum-enhanced sensing.

It is known, for instance, that \textit{criticality} is a powerful resource for metrology \cite{zanardi2008quantum,PhysRevA.88.063609}, as it allows for \textit{super-extensive} scaling of the precision in the estimation of Hamiltonian parameters, and external magnetic fields.
Similarly at finite temperatures, quantum phase crossovers allow for better metrological bounds, even when the parameter to be estimated is the temperature itself \cite{invernizzi2008optimal,salvatori2014quantum,mehboudi2015ultracold,dePascuale2015local,guo2015improved}. Quantum many-body systems exhibiting phase transitions could thus make very precise magnetometers or thermometers, if tuned close to a critical point. On the other hand, non-linearities giving rise to $ k $--body interactions between all $ N $ constituents of the probe may allow for precisions scaling like $ \delta\lambda\sim 1/N^{k-\frac12} $ without requiring any entanglement \cite{luis2004nonlinear,roy2008exponentially,boixo2007generalized}. Note that this implies that even \textit{super-Heisenberg} resolutions are achievable \cite{boixo2008quantum,woolley2008nonlinear,choi2008bose,napolitano2011interaction}. 

In this paper we focus precisely on parameter estimation of many-body Hamiltonians. Aiming at guiding practical experimental situations, we analyze equilibrium systems at finite temperature and try to answer to some relevant questions. Namely, what are the fundamental bounds on the precision of the estimation? Are there experimentally feasible measurement strategies that can saturate those bounds? Is it possible to beat the shot-noise limit at finite temperatures? Our results show that this is indeed the case. 

Specifically, we study high precision \textit{magnetometry} with a many-body probe, modeled by the paradigmatic $ XX $ Hamiltonian \cite{lieb2004two} in a thermal state. In the best-case scenario, we find that the \textit{sensitivity} $ (\delta h)^{-2} $ of the estimation of the external magnetic field $ h $ scales extensively (i.e. linearly in the `probe size' $ N $). We also find that the optimal estimator is simply the component of the total magnetization in the direction of the external magnetic field. 

An iterative feedforward scheme, allowing for the active exploitation of criticality, is then proposed. At each step of our protocol, we measure the probe and refine our estimate for $ h $. This is then used to update the internal coupling $ J $ of the many-body Hamiltonian, so that, once re-thermalized, the probe comes `closer' to the critical region. Iterating the procedure can reduce $ \delta h $ down to the level of the thermal fluctuations. We show that such an adaptive scheme exhibits a super-extensive scaling of $ \delta h \sim 1/N^{2/3} $, thus beating the shot-noise limit by a factor of $ 1/N^{1/6} $. We consider as well, the estimation of the Hamiltonian parameter $ J $.

More generally, we discuss the problem of parameter estimation in the thermal states $ \hat{\tau} = Z^{-1}\exp{(-\beta\hat{H})} $, for Hamiltonians of the form $ \hat{H} = \lambda_1\hat{H_1} + \lambda_2\hat{H_2} $ with $ [\hat{H}_1,\hat{H}_2 ] = 0 $, of which the $ XX $ model is a case study. We find that $ \hat{H}_i $ is an optimal estimator for $ \lambda_i $ (with $ i = \{1,2\} $), and that it is capable of operating at sensitivities limited by the corresponding thermodynamic susceptibility $ \chi_{\lambda_i} \equiv \partial^2\log{Z}/\partial \lambda_i^2 $. For instance, the sensitivity for magnetometry is upper-bounded by the adiabatic magnetic susceptibility $ \chi_h $. This observation highlights the interesting links between thermodynamics and estimation theory. Finally, the more general case of quantum many-body Hamiltonians with non-commuting terms (i.e. $ [\hat{H}_1,\hat{H}_2] \neq 0 $) is also studied by resorting to the \textit{XY} model, which `interpolates' between the \textit{XX} model and the Ising model \cite{zanardi2008quantum,invernizzi2008optimal,skotiniotis2015quantum}.

The paper is structured as follows: In Sec.~\ref{sec:thermal_est_intro}, we briefly revisit the basics of quantum parameter estimation in thermal states and discuss the relation between sensitivity and thermodynamic susceptibility. Next, in Sec.~\ref{sec:thermal_magnetometry}, we introduce the $ XX $ model and assess its ultimate magnetic sensitivity at finite temperatures. In Sec.~\ref{sec:sub-shot-noise}, we show how sub-shot-noise magnetometry can be achieved by exploiting criticality in our adaptive estimation scheme. Also the measurement of $ J $ is briefly considered in Sec.~\ref{sec:J_estimation}. In Sec.~\ref{sec:ising}, we consider magnetic field sensing beyond the $ XX $ model, allowing for non-commutativity between the two main Hamiltonian terms $ H_1 $ and $ H_2 $, Finally, we summarize and conclude in Sec.~\ref{sec:conclusion}.

%
 
\section{Thermal parameter estimation} \label{sec:thermal_est_intro}

As already advanced, in order to estimate an unknown parameter $ \lambda $ one has to couple it to a probe. After the interaction has taken place, the state of the probe $ \hat{\varrho}(\lambda) $ may be interrogated by performing a projective measurement onto the eigenbasis of some suitable observable $ \hat{O} $, which allows to build an estimate based on the measurement results. In order to reduce the error in the estimation, one can simply repeat the procedure $ M $ times. The ensuing statistical uncertainty can be cast as
\begin{equation}
\delta \lambda \equiv \frac{\Delta\hat{O}}{M\left|\partial_{\lambda}\langle\hat{O}\rangle_{\hat{\varrho}}\right|},
\label{eq:error}
\end{equation}
where $\Delta\hat{O} \equiv \sqrt{\langle \hat{O}^2 \rangle - \langle \hat{O} \rangle^2} $ and $ \langle\hat{O}\rangle \equiv \text{tr}\{ \hat{O}\,\hat{\varrho}(\lambda) \}$. The error $ \delta\lambda $ is lower bounded by the quantum Cram\'{e}r-Rao bound \cite{PhysRevLett.72.3439}
\begin{align}
\delta \lambda \geq\frac{1}{M\sqrt{\mathcal{F}({\lambda})}}.
\label{eq:ERROR-QCRB}
\end{align}
Here, $\mathcal{F}({\lambda})$ stands for the quantum Fisher information (QFI) associated to the parameter ${\lambda}$. In turn, the QFI is defined by
\begin{align}
\mathcal{F}({\lambda})\equiv\mathrm{tr}\left\lbrace\hat\varrho({\lambda})\,\hat\Lambda_{\lambda}^2\right\rbrace,
\label{eq:qfi}
\end{align}
where the Hermitian operator $\hat{\Lambda}_{\lambda}$ is termed `symmetric logarithmic derivative' (SLD) and stems from the relation
\begin{equation}
 \hat{\Lambda}_{\lambda}\hat{\varrho}({\lambda})+\hat{\varrho}({\lambda})\hat{\Lambda}_{\lambda}\equiv 2\partial_{\lambda}\hat{\varrho}({\lambda}).
 \label{eq:sld}
\end{equation}

When $ \hat{O} $ happens to be diagonal in the eigenbasis of $ \hat{\Lambda}_\lambda $, the inequality in Eq.~\eqref{eq:ERROR-QCRB} is saturated. That is, the SLD characterizes the most informative measurements about $ \lambda $. It is worth to mention that, measuring the QFI itself is an important and challenging task, since it is a witness of multipartite entanglement among the individual parts of a quantum (many body) system \cite{Hauke_Nat_phys_2016,Strobel_Science_2014,PhysRevA.85.022322,PhysRevA.85.022321}. In what follows we will set $ M = 1 $ (i.e. assume a single shot scenario) unless stated otherwise. Furthermore we use the notation $ F(\lambda;\hat{O}) $ to refer to the $ \lambda $-sensitivity $ (\delta\lambda)^{-2} $ of $ \hat{O} $, so that $ F(\lambda;\hat{O}) \leq \mathcal{F}(\lambda) = \sup_{\hat{O}}{(\delta\lambda)^{-2}} $. 

Unfortunately, the SLD projections can be highly non-local and hence, very hard to implement, especially on large multipartite probes. Even finding $ \hat{\Lambda}_\lambda $ analytically is sometimes a challenging task \cite{escher2011noisy,demkowicz2012elusive,alipour2014quantum}. Therefore, in practice, one must consider more manageable sub-optimal estimators, which nonetheless allow for a comparatively small error. 

Although at finite temperature quantum phase transitions become phase crossovers, some quantum effects related to criticality at $T=0$ survive at finite (but low) temperatures. Here we consider thermal states of quantum spin chains and analyze the influence of criticalilty in parameter estimation. In particular, we consider $ N $-body Hamiltonians of the form $ \hat{H} = \lambda_1\hat{H}_1 + \lambda_2 \hat{H}_2 $ as it is often the case in paradigmatic spin models. Interestingly, in the special case in which the two terms commute (i.e. $ [ \hat{H}_1, \hat{H}_2 ] = 0 $), the corresponding optimal estimator for either of the Hamiltonian parameters $ \lambda_i $ ($ i\in\{1,2\} $) and its sensitivity, may be easily found from Eqs.~\eqref{eq:qfi} and \eqref{eq:sld}. One only needs to replace $ \hat{\varrho} $ with the thermal state $ \hat{\tau} \equiv Z^{-1}\exp{(-\beta\hat{H})} $, where $ Z $ is the partition function, $ \beta \equiv (k_B T)^{-1} $, and $ k_B $ is the Botlzmann constant, and use $ \exp{(-\beta\hat{H})} = \exp{(-\beta\lambda_1\hat{H}_1)} \exp{(-\beta\lambda_2\hat{H}_2)} $. This yields 
\begin{equation}
\hat\Lambda_{\lambda_i}\hat\tau+\hat\tau\hat\Lambda_{\lambda_i} = -\beta\left(\hat H_i-\langle\hat H_i\rangle\right)\hat\tau-\beta\,\hat\tau\left(\hat{H_i}-\langle\hat H_i\rangle\right),
\label{sld-proof}
\end{equation}
implying that $ \hat{H}_i $ is itself an optimal estimator for $ \lambda_i $. According to Eq.~\eqref{eq:qfi} the corresponding sensitivity is just
\begin{equation}
\mathcal{F}(\lambda_i)=\beta^2{\Delta\hat{H}_i}^2.
\label{eq:Fisher-Variance}
\end{equation}

Making use of Eqs.~\eqref{eq:ERROR-QCRB} and \eqref{eq:Fisher-Variance} one may write the uncertainty-type relation $ {\Delta\hat{H_i}}^2\delta\lambda_i^2 \geq \beta^{-2} $. Also, given the definition of $ \delta\lambda_i $ in Eq.~\eqref{eq:error}, and using the fact that $ \langle \hat{H}_i \rangle = -\beta^{-1}\partial_{\lambda_i}\log{Z} $, the maximum $ \lambda_i $-sensitivity can be alternatively expressed as
\begin{equation}
\mathcal{F}(\lambda_i) = \beta\left\vert\frac{\partial\langle\hat{H}_i\rangle}{\partial\lambda_i}\right\vert = \frac{\partial^2\log{Z}}{\partial\lambda_i^2} = -\beta \frac{\partial^2 A}{\partial\lambda_i^2},
\label{eq:susceptibility_bound}
\end{equation} 
where $ A \equiv - k_B T \log{Z} $ stands for the Helmholtz free energy. As it can be seen, the ultimate precision in the estimation of Hamiltonian parameters from thermal states is nothing but a generalized thermodynamic susceptibility. For instance, in the case of thermometry, the specific heat is the relevant figure of merit and, as we shall see below, what limits the sensitivity of a magnetometer is its adiabatic magnetic susceptibility \cite{Hauke_Nat_phys_2016,zanardi2008quantum,PhysRevLett.114.220405,dePascuale2015local,
PhysRevA.88.063609,PhysRevE.76.022101,PhysRevLett.99.095701,PhysRevA.75.032109}.

%

\section{Magnetometry in the $ XX $ model}\label{sec:thermal_magnetometry}

\begin{figure}
\includegraphics[width=0.9\linewidth]{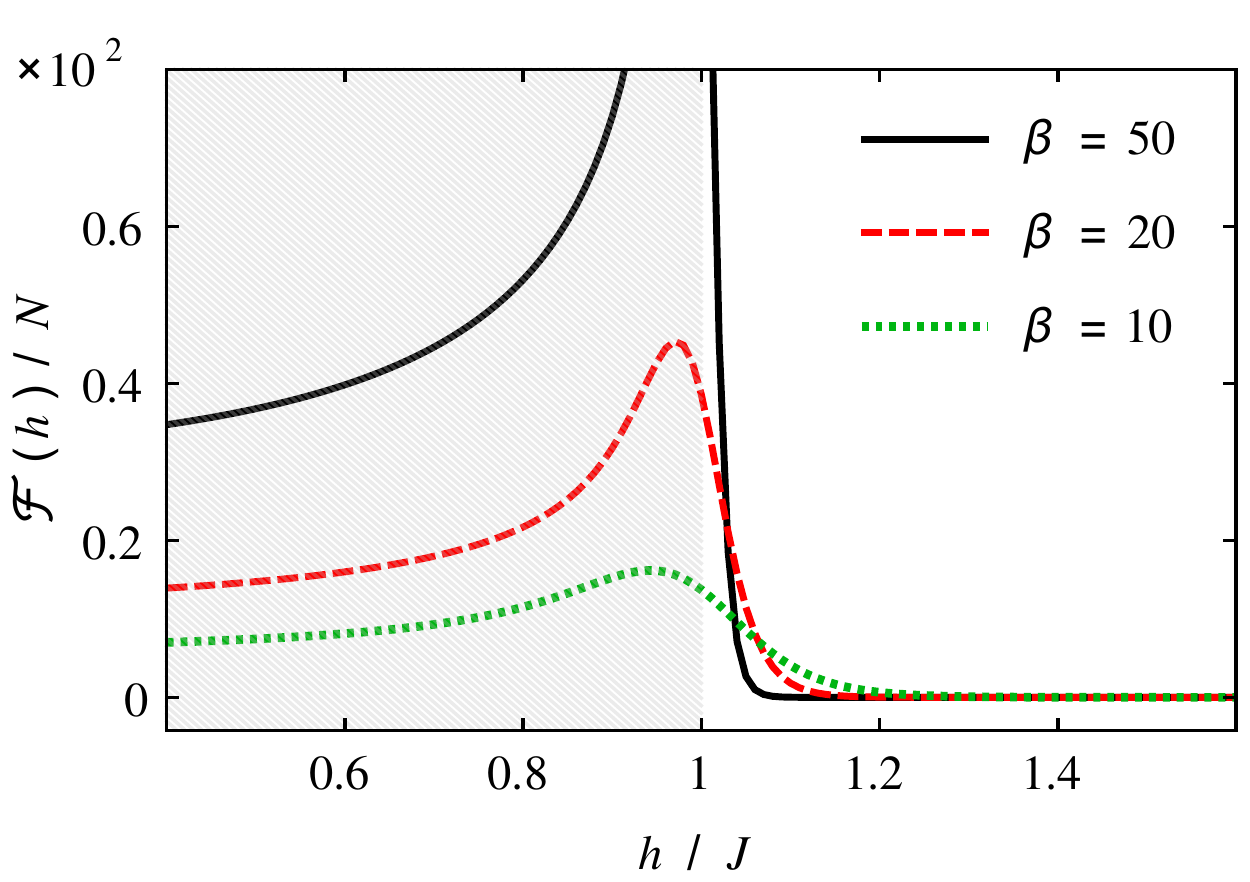}
\caption{(color online) \textit{Specific} QFI (i.e. $ \mathcal{F}(h)/N $) for the estimation of the magnetic field $ h $ in the $ XX $ model as a function of $ h/J $, at three different temperatures. The shaded area corresponds to the ferromagnetic region. In the plot $ N = 10^5 $ and $ J=1 $.} 
\label{fig1}
\end{figure}

\subsection{The one-dimensional $ XX $ Hamiltonian} \label{sec:xx_model}

Before moving on to look into the magnetic sensitivity in the \textit{XX} model, we will briefly discuss the corresponding Hamiltonian and comment on its exact solution. The \textit{XX} model is a special instance of the more general $ XY $ Hamiltonian \cite{lieb2004two}. It encompasses $ N $ spin 1/2 particles, arranged in a one-dimensional chain, with nearest-neighbour interactions of uniform strength $ J $ along the $ x $ and $ y $ directions, together with a transverse magnetic field of intensity $ h $ along the $ z $ axis. The $ XX $ Hamiltonian writes as   
\begin{eqnarray}
\hat{H}_{XX} = -\frac{J}{2}\sum_{i=1}^{N}\left( \hat\sigma^{(x)}_{i}\hat\sigma^{(x)}_{i+1} + \hat\sigma^{(y)}_{i}\hat\sigma^{(y)}_{i+1}\right)-h\sum_{i=1}^N\hat\sigma_{i}^{(z)},
\label{eq:XXmodel-nondiag}
\end{eqnarray}
where $\hat\sigma_i^{(\alpha)} $ stands for the $ x $, $ y $ or $ z $ Pauli operators of the spin in the $ i $-th site. For simplicity, we assume even $ N $ and periodic boundary conditions $ \hat{\sigma}^{(\alpha)}_{N+1} = \hat\sigma_1^{(\alpha)} $, although these details have little impact on the results, for large enough $ N $.

At zero temperature and in the thermodynamical limit, this system has two magnetic phases: For $\vert h \vert \gg \vert J \vert$, the ground state of $ \hat{H}_{XX} $ becomes a tensor product, with all spins `aligned' along the magnetic field; this is the paramagnetic phase. In the opposite limit of $\vert h \vert \ll \vert J \vert $ one finds a degenerate ferromagnetic (anti-ferromagnetic) ground state for $ J > 0$ ($ J < 0 $). At the critical point $ \vert h \vert = \vert J \vert $, the system undergoes a second-order phase transition. In what follows, we assume $ J > 0 $. It must be noted that, strictly speaking, this is so only at zero temperature. Signatures of quantum criticality are, however, still detectable at finite temperatures and can be exploited for metrological purposes \cite{zanardi2008quantum}. 

The $ XX $ Hamiltonian can be effectively mapped into a collection of non-interacting fermions via the Jordan-Wigner and Fourier transformations \cite{lieb2004two, mikeska2004one}. This yields 
\begin{align}
\hat H_{XX} = \sum \epsilon_p \hat\alpha_p^{\dagger} \hat\alpha_p.
\label{eq:XXmodel-diag}
\end{align}

Here, $\hat\alpha_p$ ($ \hat\alpha_p^{\dagger} $) is an annihilation (creation) operator corresponding to a fermionic mode $ p $ of energy
\begin{align}
\epsilon_p &= 2J\,(\cos{p}-h/J),\nonumber\\
p &= \frac{\pi}{N}(2l+1),\qquad l \in\{-N/2,\cdots,N/2-1\}.
\label{eq:energies_xx_fermions}
\end{align}

Simple spin Hamiltonians like the $ XX $ model can be experimentally simulated with ultracold atomic gases trapped in optical lattices \cite{lewenstein2007ultracold}. Furthermore, the state of such effective spin models can be probed by resorting to a non-demolition scheme based on quantum polarization spectroscopy \cite{kupriyanov2005multimode}: The angular momentum of the sample may be coupled to the polarization of an incident laser beam and read out by homodyne detection of the scattered light \cite{eckert2008quantum,eckert2007quantum,de2011probing}. 

\subsection{Optimal measurement and its magnetic sensitivity}\label{sec:optimal_est_h}

It is easy to see that the terms proportional to $ J $ and $ h $ in Eq.~\eqref{eq:XXmodel-nondiag} commute with each other. Hence, from Sec.~\ref{sec:thermal_magnetometry}, we know that the optimal observable for the estimation of $ h $ is the total magnetization in the $ z $-direction $ \hat{J}_z = \sum_{i=1}^N \hat{\sigma}_i^{(z)} $ and that the corresponding sensitivity is modulated by the magnetic susceptibility $ \mathcal{F}(h) = \beta\,\vert\partial_h\langle\hat{J}_z\rangle\vert \equiv \beta\,\chi_h $ \cite{quan2009finite}. Note that $ \langle\hat{J}_z\rangle $ is precisely the kind of quantity that can be accessed experimentally using quantum polarization spectroscopy.

Eqs.~\eqref{eq:XXmodel-diag} and \eqref{eq:energies_xx_fermions} allow to write the magnetization in the thermal state $ \hat{\tau} $ as
\begin{align}
\langle\hat{J}_z\rangle=2\sum_pn_p-N,
\label{eq:XX-mean-Jz}
\end{align}
with $ n_p = [1+\exp{(\beta\epsilon_p)}]^{-1} $ being the fermionic thermal occupation number of the $p$-th energy level. The explicit formula for $ \mathcal{F}(h) $ follows from Eq.~\eqref{eq:susceptibility_bound}, which yields
\begin{align}
\mathcal{F}(h)=\beta\left\vert\frac{\partial \langle\hat{J}_z\rangle}{\partial h}\right\vert = 4\,\beta^2\sum_pn_p(1-n_p).
\label{eq:XX-QFI}
\end{align}

As shown in Fig.~\ref{fig1}, the sensitivity peaks in the ferromagnetic phase, close to the critical point $ \vert h/J \vert = 1 $. This feature becomes sharper as the probe is cooled down, until $ \mathcal{F}(h) $ eventually diverges at criticality in the limit $ \beta \rightarrow \infty $ \cite{zanardi2008quantum}. Note as well that $ \mathcal{F}(h) $ drops quickly to zero as the probe enters the paramagnetic region (most markedly at low temperatures), whereas it remains non-vanishing within the ferromagnetic phase. This is intuitive, recalling that the paramagnetic ground state is an eigenstate of the estimator $ \hat{J}_z $ and thus, completely insensitive to fluctuations in the field intensity $ h $. 

Interestingly, although increasing the equilibrium temperature of the probe significantly reduces the attainable sensitivity both in the ferromagnetic phase and at criticality, thermal mixing does slightly enhance the sensitivity of paramagnetic samples. This is not so surprising, as an increase in the temperature of the sample populates excited states of $ \hat{H}_{XX} $ (more sensitive than the paramagnetic ground state), albeit at the expense of the overall destruction of quantum correlations.

\section{Sub-shot-noise sensing in the \textit{XX} model}\label{sec:sub-shot-noise}

\subsection{Low-temperature approximation for $ \mathcal{F}(h) $}\label{sec:low_temp_approx}

In what follows, it will be useful to have a simple working approximation for $ \mathcal{F}(h) $ in the paramagnetic phase, capturing its dependence on $ N $, $ h/J $, and $ \beta $. Specifically, we are interested in the regime of low temperatures ($ \beta\gg 1 $) and large $ N $. Looking at Eq.~\eqref{eq:XX-QFI}, we see that the contribution of terms with $ n_p \simeq \{0,1\} $ to $ \mathcal{F}(h) $ can be safely neglected. Recalling that $ n_p = [1+\exp{(\beta\epsilon_p)}]^{-1} $, only those terms for which $ \beta\,\vert\epsilon_p\vert < \kappa $, where $ \kappa $ is some small positive constant, contribute significantly to the total magnetic sensitivity. From Eq.~\eqref{eq:energies_xx_fermions} it follows that $ -\kappa/\beta < 2J(\cos{p}-h/J) < \kappa/\beta $, and hence
\begin{equation}
\arccos{\left(\frac{h}{J} - \frac{\kappa}{2\beta J}\right)} < p < \arccos{\left(\frac{h}{J} + \frac{\kappa}{2\beta J}\right)}. 
\label{eq:scaling_ineq}
\end{equation}
One can now perform Taylor expansion to first order in the small parameter $ \kappa/\beta $, which yields
\begin{equation}
\arccos{\left(\frac{h}{J} \pm \frac{\kappa}{2\beta J}\right)} \simeq \arccos{(h/J)} \mp \frac{\kappa}{2\beta J\sqrt{1-(h/J)^2}}.
\label{eq:scaling_taylor}
\end{equation}

Since $ N $ is large, we may assume that the indices $ p $ are continuously distributed with a uniform `density' $ N / 2\pi $ (recall that $ \vert p \vert < \pi $). Therefore, the number of energy levels effectively contributing to the sum in Eq.~\eqref{eq:XX-QFI} would be the product of $ N/2\pi $ and the gap between the upper and lower bounds of Eq.~\eqref{eq:scaling_ineq}. Taking a constant $ n_p $ for all the terms involved in the sum gives an optimal magnetic sensitivity of  
\begin{align}
\mathcal{F}(h)\simeq\mathcal{F}_\text{app}(h) \equiv C \frac{\beta N }{J\sqrt{1-(h/J)^2}}.
\label{eq:qfi_approximation}
\end{align}

\begin{figure}
\includegraphics[width=0.9\linewidth]{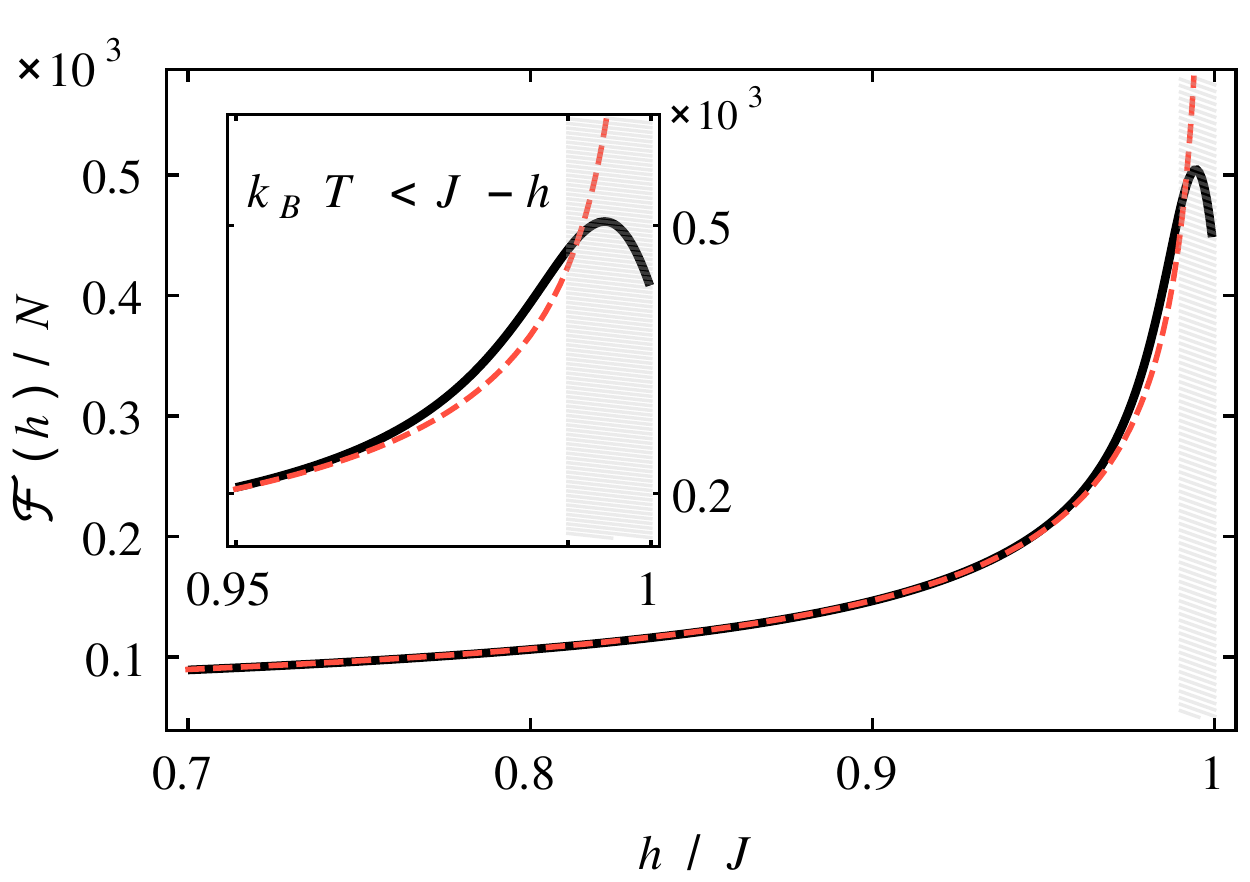}
\caption{(color online) (solid black) Specific QFI for magnetic field sensing in the $ XX $ model versus $ h/J $. All the plotted area lies within the ferromagnetic phase. (dashed red) Low-temperature approximation of the magnetic sensitivity $ \mathcal{F}_\text{app}(h) $ from Eq.~\eqref{eq:qfi_approximation}. The region $ k_B T > J-h  $ where the approximation breaks down appears in shaded gray. (inset) Close-up of the neighbourhood of the critical point. The temperature was set to $ \beta = 100 $, $ N = 10^{4} $, and $ J = 1 $.} 
\label{fig2}
\end{figure}

For low enough temperatures and large $ N $, we numerically find the fitting parameter $ C \approx 0.64 $ to be independent of $ \beta $, $ J $, and, most importantly, also of the size of the probe $ N $. The good agreement between Eq.~\eqref{eq:qfi_approximation} and $ \mathcal{F}(h) $ is showcased in Fig.~\ref{fig2}. As a rule of thumb, we can expect the approximation to hold so long as $ k_B T < J-h $. Closer to the critical point, i.e. when $ J-k_B T < h < J $, the magnetic sensitivity presents a maximum of approximately $ \mathcal{F}_\text{app}(h = J-k_B T) $. 

Finally, notice as well that $ \mathcal{F}_\text{app}(h) $ is linear in $ N $ or, in other words, the magnetic sensitivity scales extensively with the probe size. Next, we will show how the scaling of $ (\delta h)^{-2} $ may be enhanced by means of a feedforward adaptive protocol that actively exploits quantum criticality. 

\subsection{Adaptive feedforward magnetometry}\label{sec:adaptive_h}

The expression provided for $ \mathcal{F}_\text{app}(h) $ (Eq.~\eqref{eq:qfi_approximation}), suggests an adaptive protocol to improve the estimation of $ h $. Let us assume, in full generality, that $ h $ is known within an interval $ h_\text{min} < h < h_\text{max} $. If the Hamiltonian parameter $ J $ is accessible to control, one may start by tuning it to $ J = h_\text{max} $ to ensure that the spin chain lies in the ferromagnetic side of the transition. After the sample has equilibrated with the new parameters, we can measure its magnetization $ \hat{J}_z $ and come up with the estimate $ h \pm \delta h_1 $, with `error bars' $ \delta h_1 \simeq 1/\sqrt{\mathcal{F}_1} $, where
\begin{equation}
\mathcal{F}_1 \equiv C\frac{\beta\nu N}{h_\text{max}\sqrt{1-(h/h_\text{max})}}\equiv A \nu N.
\label{eq:first_iteration}
\end{equation} 

In Eq.~\eqref{eq:first_iteration}, we have explicitly accounted for enough repetitions $ \nu $ of this first step to ensure that $ \delta h_1/h \ll 1 $. At this point, one can update the interaction strength to $ J = h + \delta h_1 $ and, again after re-equilibration of the probe, refine the estimate of $ h $ according to the outcomes of $ \nu $ additional magnetization measurements. The error $ \delta h_2 $ after the second iteration is arguably much smaller than $ \delta h_1 $. Note that the protocol is essentially driving the probe towards the critical point, which drastically increases the sensitivity as shown in Fig.~\ref{fig2}. In particular, $ \delta h_2 \simeq 1/\sqrt{ \mathcal{F}_2} $, where 
\begin{equation}
\mathcal{F}_2 = \left(\frac{C\beta}{\sqrt{2 h}}\right)\frac{\nu N}{\sqrt{\delta h_1}} + \mathcal{O}\,\left(\frac{\delta h_1}{h}\right)^{3/2}\simeq B\nu N \mathcal{F}_1^{1/4},
\label{eq:second_iteration}
\end{equation}
where $ B \equiv C\beta/\sqrt{2h} $. $ \mathcal{F}_2 $ As the protocol is repeated further, we find $ \delta h_k \simeq 1/\sqrt{\mathcal{F}_k} $, with
\begin{multline}
\mathcal{F}_k \simeq B \nu N F_{k-1}^{1/4} = A^{1/4^{k-1}}B^{1+1/4+\cdots+1/4^{k-2}}(\nu N)^{1+1/4+\cdots+1/4^{k-1}} \\
= A^{1/4^{k-1}}B^{4/3(1-1/4^{k-1})}(\nu N)^{4/3(1-1/4^{k})}.
\label{eq:kth_iteration}
\end{multline}

In the limit of large $ k $, the sensitivity scales as $ F_k \sim N^{4/3} $ so that $ \delta h_k \sim 1/ N^{2/3} $, which outperforms the shot-noise scaling by a factor of $ 1/ N^{1/6} $. Hence, the proposed adaptive scheme shows that, at finite temperatures, it is possible to exploit criticality in its wider sense to allow quantum-enhanced sensing and overcome the linear (shot-noise) scaling associated to uncorrelated probes. The reason for such (certainly surprising) fact is that, at each step $ k $ of the protocol, the thermal state changes and approaches the quantum crossover point, with its critical behavior. Those changes reflect in a sensitivity that scales super-extensively with the number of particles $ N $. This is the main result of the work presented here. 

Two clarifications are in order. To begin with, note that for Eq.~\eqref{eq:qfi_approximation} to remain applicable, we shall always work in the limit $ \{ \beta, N \} \gg 1 $. Recall, however, that the approximation $ \mathcal{F}_\text{app}(h) $ only holds if $ k_B T < J_k - h = \delta h_{k-1} $, so that thermal fluctuations set an effective lower bound for the statistical uncertainty attainable with this iterative scheme: As soon as $ \delta h $ falls below $ k_B T $, updating the interaction strength provides no scaling advantage. Indeed, it may even become detrimental if the probe is pushed too close to criticality (see Fig.~\ref{fig2}). Note that this does not mean that uncertainties $ \delta h $ below the level of thermal fluctuations are unattainable, but only that the error decreases no faster than $ 1/N^{1/2} $ beyond that point.

Secondly, the only metrologically relevant resource considered in our analysis is the number $ N $ of spins in the sample. In particular we implicitly assume that the precise adjustment of $ J $, the iteration of the magnetization measurement $ \nu\times k $ times, or the re-thermalization of the probe at each step come at no additional cost. Care must be taken, however, as this may be not the case in actual experiments: Practical limitations, like the short lifetime of the sample or the imperfect control of the Hamiltonian parameters, may call for a different assessment of resources, specific to each particular implementation.     

\subsection{Sub-shot-noise estimation of the coupling $ J $}\label{sec:J_estimation}

\begin{figure}
\includegraphics[width=0.9\linewidth]{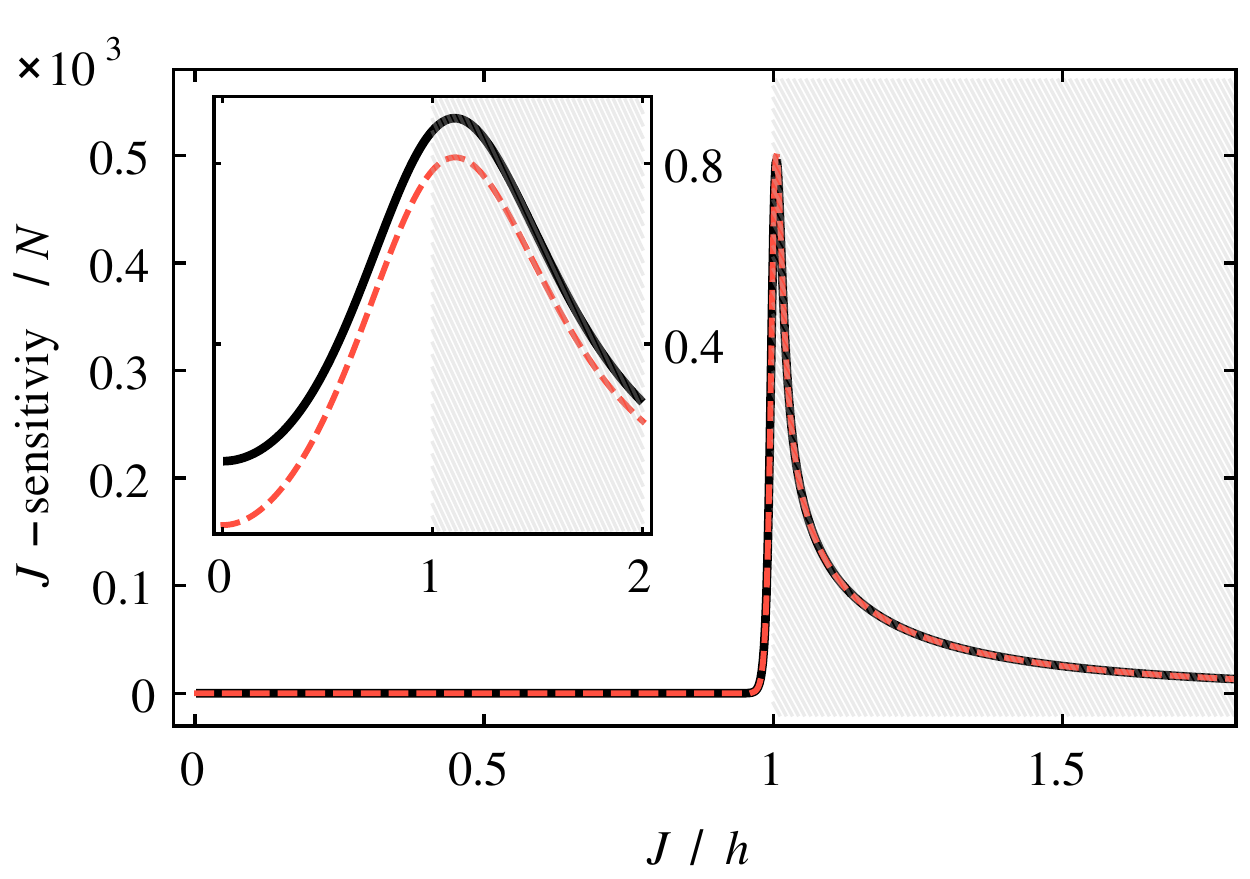}
\caption{(color online) (solid black) Specific QFI for the estimation of $ J $ in the $ XX $ model versus $ J/h $. As in Fig.~\ref{fig1} the ferromagnetic phase has been shaded. (dashed red) Specific $ J $-sensitivity $ F(J;\hat{J}_z)/N $ of the total magnetization $ \hat{J}_z $. The temperature was set to $ \beta = 100 $, $ h = 1 $, and $ N = 10^{3} $. (inset) Same as in the main plot for the much larger temperature $ \beta = 2 $. Note that, unlike in the main plot, the vertical axis of the inset is not scaled by the $ 10^3 $ factor.} 
\label{fig3}
\end{figure}

For completeness we also address the estimation of the Hamiltonian parameter $ J $ in the $ XX $ model. As we know from Sec.~\ref{sec:thermal_est_intro}, the estimator $ \hat{O}_J \equiv \sum_{i=1}^N \left(\hat{\sigma}_i^{(x)}\hat{\sigma}_{i+1}^{(x)} + \hat{\sigma}_i^{(y)}\hat{\sigma}_{i+1}^{(y)}\right) $ would be optimal in this case. Its sensitivity can be obtained as in Eq.~\eqref{eq:XX-QFI}, which gives 
\begin{align}
\mathcal{F}(J)=4\beta^2\sum_p{\cos^2{p}~n_p(1-n_p)}.
\label{eq:XX_QFI_J}
\end{align}

Unfortunately, $ \hat{O}_J $ is not as easy to measure as the total magnetization $ \hat{J}_z $, since it involves two-body correlations. Although generally sub-optimal, the magnetization is known to be a good estimator for $ J $ in the related Ising model (cf. Sec.~\ref{sec:ising}) \cite{invernizzi2008optimal}, which motivates us to look at the $ J $-sensitivity $ F(J;\hat{J}_z) $ of this observable. This is plotted alongside $ \mathcal{F}(J) $ in Fig.~\ref{fig3}. Note that the abscissa is, in this case, $ J/h $ instead of $ h/J $. As in Fig.~\ref{fig1}, $ \mathcal{F}(J) $ (solid black) peaks in the ferromagnetic phase close to the critical point. On the other hand, $ F(J;\hat{J}_z) $ (dashed red) is seen to be nearly optimal at low enough temperatures. Most interestingly,  $ F(J;\hat{J}_z) $ still remains very close to the optimal sensitivity even at very large temperatures, as illustrated in the inset of Fig.~\ref{fig3}. Hence, $ \hat{J}_z $ can be regarded as a practical alternative for estimating $ J $.

Due to the similarity between Eqs.~\eqref{eq:XX-QFI} and \eqref{eq:XX_QFI_J}, one may proceed as in Sec.~\ref{sec:low_temp_approx} to come up with the following low-temperature approximation for $ \mathcal{F}(J) $ at large $ N $:
\begin{equation}
\mathcal{F}(J)\simeq\mathcal{F}_\text{app}(J)\equiv C\frac{h^2 \beta N}{J^3\sqrt{1-(h/J)^2}}.
\label{eq:qfi_approximation_J}
\end{equation}

Consequently, the exact same line of reasoning of Sec.~\ref{sec:adaptive_h} applies to this case: Iteratively updating the value of the external magnetic field $ h $, so as to drive the system towards the critical point, allows, in principle, for sub-shot-noise scaling in the $ J $-sensitivity.  


\section{Magnetometry beyond the \textit{XX} model}\label{sec:ising}

\begin{figure*}
\includegraphics[width=0.48\linewidth]{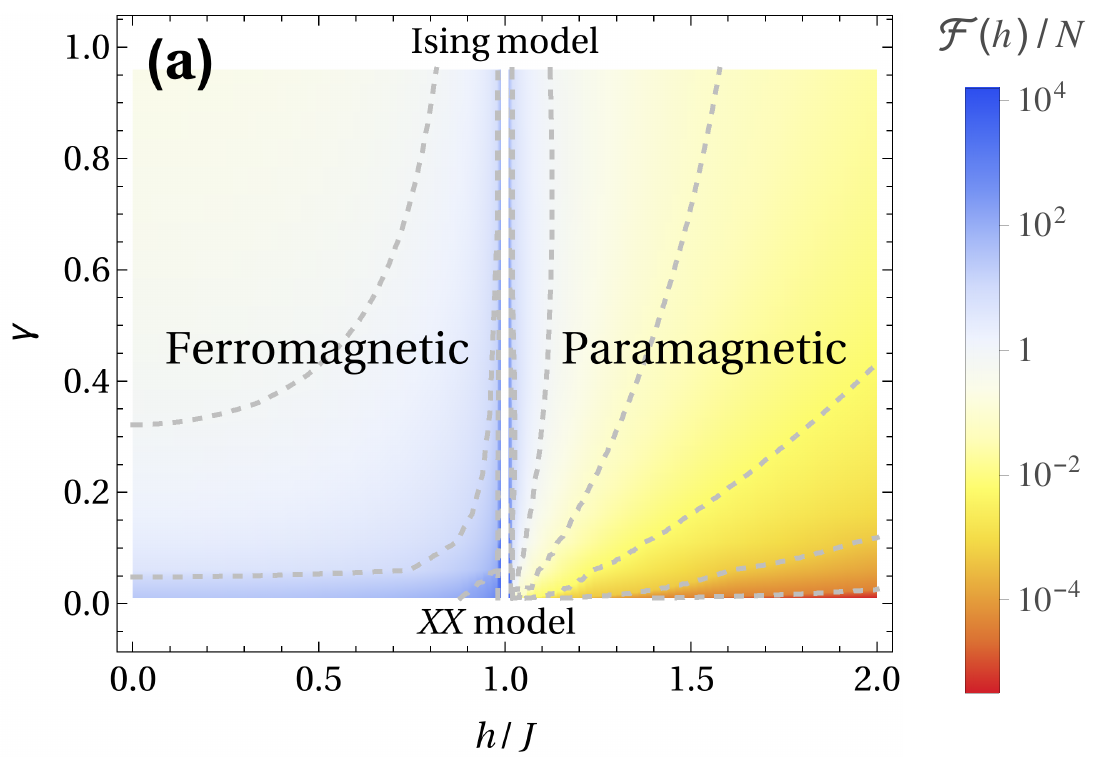}
\includegraphics[width=0.43\linewidth]{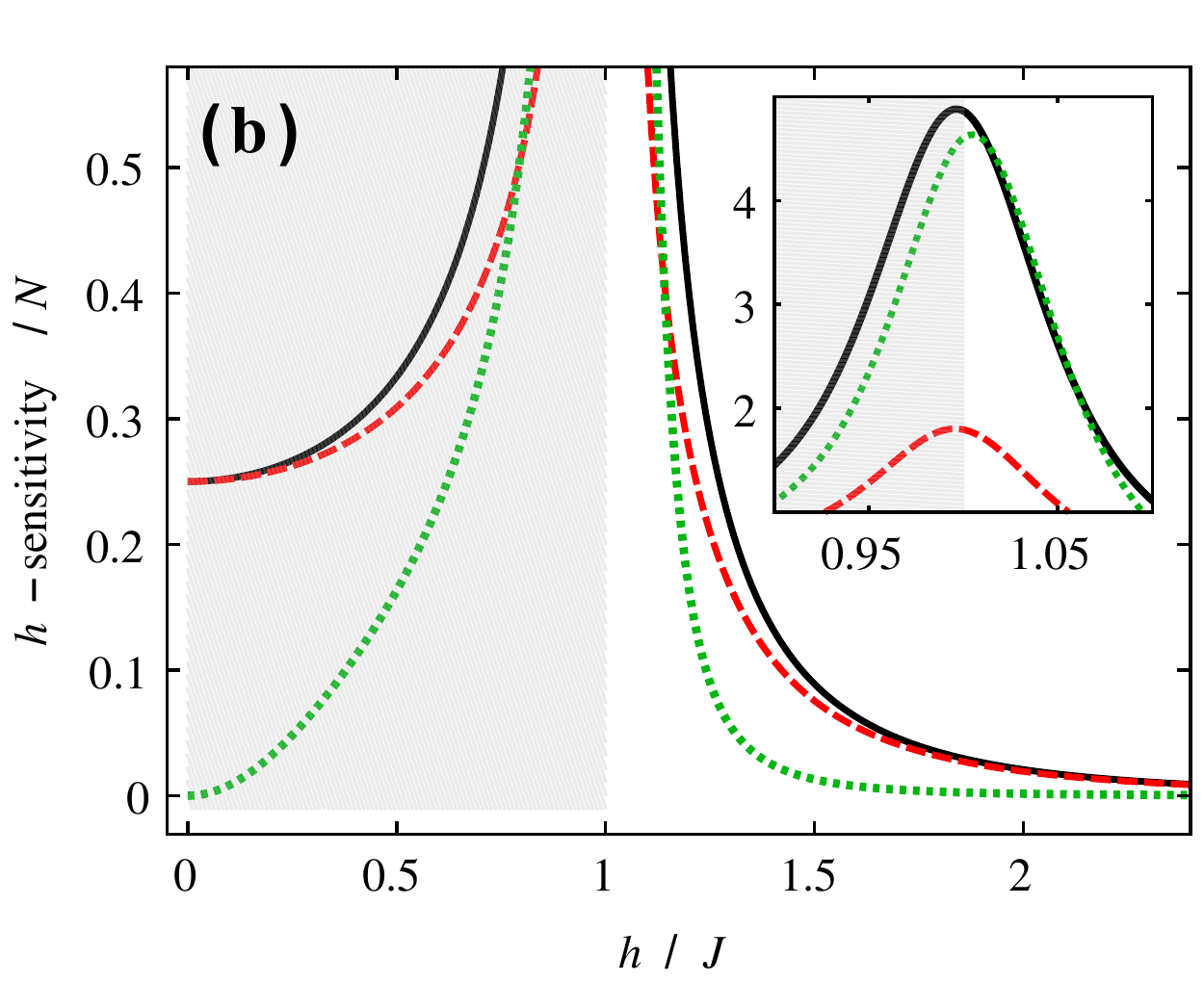}
\caption{(color online) (a) Specific QFI (in logarithmic scale) in the $ XY $ model as a function of $ h/J $ and $ \gamma $. The critical line appears highlighted in white. Note that $ \gamma = 0 $ corresponds to the $ XX $ model and $ \gamma = 1 $, to the Ising model. The sensitivity increases with the asymmetry parameter $ \gamma $ in the paramagnetic region, whereas it decreases with $ \gamma $ in the ferromagnetic phase. For this illustration $ N = 10^3 $, $ \beta = 10^3 $, and $ J = 1 $. (b) (solid black) Specific QFI in the Ising model, (dashed red) specific $ h $-sensitivity $ F(h;\hat{J}_z)/N $ of the total magnetization in the $ z $-direction, and (dotted green) specific $ h $-sensitivity $ F(h;\hat{J}_x^2)/N $. (inset) Zoom into the high-sensitivity region, not shown in the main plot. The ferromagnetic phase is hihglighted in shaded gray and $ N=40 $, $ \beta = 100 $, and $ J=1 $.} 
\label{fig4}	
\end{figure*}

We shall now turn our attention to the Heisenberg $ XY $ model, which includes the $ XX $ model as a particular case \cite{lieb2004two}. Its Hamiltonian $ \hat{H}_{XY} $ takes the form
\begin{equation}
\hat H_{XY} = -J\sum_{i=1}^{N}\left( \frac{1+\gamma}{2}\hat\sigma^{(x)}_{i}\hat\sigma^{(x)}_{i+1} +\frac{1-\gamma}{2}\hat\sigma^{(y)}_{i}\hat\sigma^{(y)}_{i+1}\right) - h\sum_{i=1}^N\hat{\sigma}_i^{(z)},
\label{eq:xy_hamiltonian}
\end{equation}
where the \textit{asymmetry parameter} $ \gamma \in [0,1] $ allows to interpolate between the $ XX $ and the Ising Hamiltonian.
The $ XY $ model can also be analytically solved with the same procedure as the $ XX $ model. After performing the Jordan-Wigner transformation followed by Fourier and Bogoliubov transformations one moves into the free fermionic representation, with the resulting energy spectrum
\begin{align}
\epsilon_p &= 2J \sqrt{\left(\cos{p}-\frac{h}{J}\right)^2+\left(\gamma\sin{p}\right)^2},\nonumber\\
p &= \frac{\pi}{N}(2l+1),\qquad l \in\{-N/2,\cdots,N/2-1\}.
\label{eq:energies_xy_fermions}
\end{align}

The maximum magnetic sensitivity of a thermal state $ \hat\tau $ of the $ XY $ Hamiltonian can be computed easily by noticing that
\begin{equation}
\hat{\tau} = \bigotimes_p \frac{\vert 0_p\rangle\langle 0_p \vert + e^{-\beta\epsilon_p} \vert 1_p \rangle\langle 1_p \vert}{1+e^{-\beta\epsilon_p}} \equiv \bigotimes_p \hat\tau_p,
\label{eq:thermal_product}
\end{equation}
where $ \vert 0_p \rangle $ ($ \vert 1_p \rangle $) denotes the empty (occupied) $ p $-th state. Using the fact that the QFI is additive under tensor products, the total magnetic sensitivity reduces to the sum of individual contributions from thermalized two-level systems $ \mathcal{F}(h) = \sum_p \mathcal{F}_p(h) $. These can be calculated as \cite{paris2009quantum,0253-6102-61-1-08}
\begin{equation}
\mathcal{F}_p(h) = 4 \sum_{i,j} \langle i_p \vert\hat\tau_p\vert i_p \rangle\left\vert \langle i_p \vert\partial_h\hat{\tau}_p\vert\ j_p\rangle \right\vert^2,
\label{eq:qfi_finite_d}
\end{equation}
where $ i,j \in \{ 0,1 \} $. When evaluating $ \partial_h\hat{\tau}_p $, one must take into account that the state vectors $ \vert i_p \rangle $ do depend on $ h $. In Fig.~\ref{fig4}(a) the resulting $ \mathcal{F}(h) $ is plotted versus $ h/J $ and $ \gamma $. Note that the sensitivity peaks sharply around the critical line $ h/J = 1 $ (indicated in white) for any $ \gamma $. Otherwise, in the ferromagnetic phase, the sensitivity decreases as the asymmetry $ \gamma $ is increased, while in the paramagnetic phase, it grows instead. 

Note that the two terms in Eq.~\eqref{eq:xy_hamiltonian} do not commute in general and, consequently, $ \hat{J}_z $ is not necessarily an optimal magnetic field estimator. Even if the QFI can be readily computed, finding the SLD is a much harder task, typically yielding complex non-local optimal estimators. It is therefore important to find good practical estimators, as in Sec.~\ref{sec:J_estimation}. 

In particular, we shall consider again $ \hat{J}_z $ and the variance $ {\Delta\hat J_x}^2 $, which can be expressed in terms of two-body correlation functions \cite{mehboudi2015ultracold}. Their corresponding $ h $-sensitivities, $ F(h;\hat{J}_z) $ and $ F(h;{\Delta\hat{J}_x}^2) $, are easy to calculate numerically for low $ N $. These are compared to $ \mathcal{F}(h) $ in Fig.~\ref{fig4}(b) for $ \gamma = 1 $ (i.e. in the Ising model). At low temperatures and far from the critical point, $ \hat{J}_z $ turns out to be nearly optimal again. In contrast, close to criticality $ {\Delta\hat{J}_x}^2 $ features a magnetic sensitivity much closer to the ultimate bound. At larger temperatures, however, correlations are destroyed by thermal mixing and, consequently, $ F(h;{\Delta\hat{J}_x}^2) $ reduces significantly. On the other hand, $ F(h;\hat{J}_z) $ remains close to optimality even at very large temperatures.
Fig.~\ref{fig4}(b) suggests that, in a practical situation, a first estimate $ h \pm \delta h_1 $ would be best obtained with the more conservative estimator $ \hat{J}_z $. If the temperature is low enough and $ J $ can be tuned to $ h + \delta h_1 $, further estimates based on $ {\Delta\hat{J}_x}^2 $ would subsequently provide much better accuracies.

\section{Conclusions}\label{sec:conclusion}
We have discussed parameter estimation in quantum spin chains at finite temperature near quantum phase crossovers. In particular, we have been concerned with magnetic field estimation in the $ XX $ model. We have shown that the corresponding sensitivity is modulated by the adiabatic magnetic susceptibility of the probe. We have also seen how, even though the magnetic susceptibility scales extensively (i.e. linearly) in the probe size, sub-shot-noise reduction of the error is still possible through a feedforward adaptive scheme. This super-extensive behaviour of the sensitivity can be maintained until the error falls below the level of the environmental noise. Additionally, we have seen that the component $ \hat{J}_z $ of the total magnetization in the direction of the external field $ h $ is strictly optimal for the estimation of $ h $ and quasi-optimal for the estimation of the internal interaction strength $ J $. Interestingly, observables like $ \hat{J}_z $ can be spectroscopically measured on spin systems, causing minimum disturbance. 

Finally, we have extended our study to more general Hamiltonians where no commutation relations can be exploited for metrology, like the paradigmatic $ XY $ model. There, the sensitivities of different sub-optimal observables have been benchmarked against the practically unattainable ultimate precision bound set by the quantum Fisher information. 

Our results may be particularly relevant to practical sub-shot-noise sensing, as we place the focus on the sensitivities achievable with probes prepared in robust thermal states, rather than the fragile highly-entangled pure preparations which are often sought in order to attain better-than-classical error scaling in parameter estimation.

The problem of the simultaneous measurement of several parameters (e.g. $ h $ and $ J $) with quantum many-body probes remains an open question that certainly deserves investigation. Although technically very challenging, it would also be interesting to extend this type of analysis to non-integrable thermal spin models, possibly featuring a richer phase diagram. This will be the subject of future work.

\subsection*{Acknowledgements}

We are thankful to Alex Monras, John Calsamiglia, Michalis Skotiniotis, and Manuele Landini for fruitful discussions. We gratefully acknowledge financial support from EU Collaborative Project TherMiQ (Grant Agreement 618074), Spanish MINECO (Project No. FIS2013-40627-P), and Generalitat de Catalunya (CIRIT Project No. 2014 SGR 966). L.A.C. acknowledges funding from the European Research Council (ERC) Starting Grant GQCOP (Grant No. 637352). M.M. and L.A.C. gratefully acknowledge support from the COST Action MP1209.  


\bibliographystyle{iopart-num}

\providecommand{\newblock}{}

\end{document}